# Love wave gas sensor based on DWNTs sensitive material

H. Hallil[1], J.L. Lachaud[1],
C. Dejous[1], D. Rebière[1],
[1]Univ. Bordeaux,
Bordeaux INP, CNRS,
IMS UMR 5218
F-33405 Talence, France
hamida.hallil@ims-bordeaux.fr

Q. Zhang[2], P. Coquet [2,3]
[2]CINTRA,
CNRS/NTU/THALES,
UMI 3288
Singapore 637553,
Singapore
[3]Univ. Lille, IEMN UMR
CNRS 8520 F-59652
Villeneuve d'Ascq,
France

E. Flahaut[4]
[4]Univ. Paul Sabatier,
CIRIMAT/LCMIE,
CNRS UMR 5085,
Toulouse, France

*Abstract*—**This work focuses on the application related to the detection of low moisture and environmental pollutants. A novel gas sensor with inkjet printed Double Walled Carbon Nano Tubes (DWNTs) on a Love wave sensor platform was developed for Volatile Organic Compounds (VOCs) and humidity detection application. The experiments were conducted in real-time at ambient conditions. Results demonstrate the adsorption of vapor compounds on DWNTs sensitive material and leads for example to frequency shifts of 1.97 kHz and 2.93 kHz with 120 ppm of ethanol vapor and 6.22 % RH, respectively.**

## I. Introduction

Air pollution caused by transport and industry has become one of major challenges. It strongly influences the health of people and the social behavior all over the world. On average, air pollution reduces life expectancy of about a year, according to the Agency for Environment and Energy Management. Chemical sensors for gas discrimination and quantification pose one of the outstanding challenges and need elaborate investigations on the interaction of the analyte gases and functional materials in terms of structural, chemical, and morphological attributes to better engineer the sensing properties of gas sensors [1, 2].

In the past few years, Carbon nanotubes (CNTs) have emerged as wonderful materials and are being considered for a wide variety of applications ranging from large scale structures to Nano electronics. There can be listed as three main types: Multiple Walled Carbon Nano Tubes (MWNTs), Double Walled Carbon Nano Tubes (DWNTs) and Single Walled Carbon Nano Tubes (SWNTs). These types of CNTs have different molecular geometries, which obviously imply different electrical and mechanical properties. For example, DWNTs and MWNTs have similar properties as SWNTs but they are much more resistant to chemicals, which is a big issue for the SWNTs. Hence, they are better suited for functionalization [3], whereas covalent functionalization of SWNTs would break some C=C double bonds, leaving holes in the structure, thus modifying electrical and mechanical properties. In the case of DWNTs, only the outer wall diameter is modified by functionalization, while the inner wall keeps similar properties as SWNTs [4]. These carbon materials are the strongest materials yet discovered in terms of tensile strength and elastic modulus. They have also a quite low density as a solid (about 1.4 g/cm²), which leads to a high specific strength of up to 48,000 kN.m.kg$^{-1}$ [5-6].

Techniques have been developed to produce high quantity of nanotubes, among them plasma arcing or laser ablation, and most of all, Chemical Vapor Deposition technique (CVD). Typical MWNTs grown by the arc-discharge method have a diameter about 20nm, while CVD grown nanotubes can have diameters up to 100nm [7]. CVD growth of CNTs can occur in vacuum or at atmospheric pressure. Major improvements in techniques have been making them more commercially available for these last 10 years.

The range of applications for CNTs is wide, from bicycle components to aero-spatial, they can be used for their exceptional strength in cables for example, or their electrical properties in Nano electronics. More specifically, they are used in this study for gas sensor applications. In that perspective, DWNTs are used as a sensitive layer to attract and immobilize different kinds of molecules like Volatile Organic Compounds (VOCs) and humidity, as pollutant and industrial toxic gases. Applications for these gas sensors might be embedded sensors for automobile industry or electronic nose for environmental applications.

For this study, the device is based on acoustic transduction associated with DWNTs printed as gas sensitive layer. This material was chosen as printing ink solution, for its remarkable mechanical properties necessary to promote the propagation of the acoustic wave and for enhanced sensitivity to target pollutants, due to high specific surface area. In a first part, the geometry and the manufacturing of the Love wave sensor and inkjet printed DWNTs are described, as well as results of electrical characterization. Then, a real time exposure to several concentrations of ethanol, toluene and humidity is used to study the sensitive material behavior. We summarize and discuss the vapors characterization results and the sensor sensitivity as proof of concept. Finally, perspectives are presented.

## II. METHODS AND MATERIALS

### A. Love Wave Sensor

The Love wave device consists of two delay lines built on an AT-cut quartz substrate. An orientation with a wave propagation direction perpendicular to the X-crystallographic axis is chosen in order to generate pure shear waves. The Love wave is generated and detected by means of interdigitated electrodes (IDTs) deposited on the substrate and prior to $SiO_2$ guiding layer, giving rise to shear horizontal surface acoustic wave (guided SH-SAW). The acoustic aperture and the center to center distance of transmitting and receiving IDTs are equal to $40\lambda$ and $210\lambda$, respectively. The acoustic energy is thus confined at the near surface to maximize the sensor sensitivity.

The geometry of the sensors plays an important role, as it affects the sensor working frequency. IDTs are composed of 44 split-finger pairs of gold and titanium (Ti/Au/Ti, total thickness about 150 nm) with a wavelength $\lambda$ (spatial periodicity) equal to 40 μm. Because of the spatial periodicity wavelength fixed by the fabrication process, devices operate at 117.5 MHz [8]. The sensitive layer is to be deposited on the upper delay line so as to compare the resultant response with the reference delay line (see Fig.1).

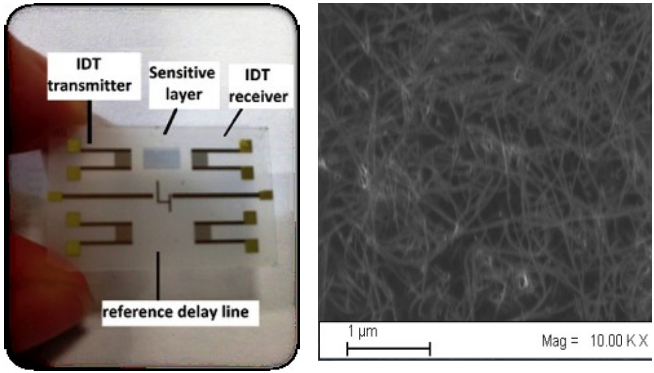

Fig. 1. (a) Inkjet-printed coated SH-SAW device with sensitive layer of DWNTs and (b) SEM image of the printed DWCNTs.

### B. Manufacturing process of the DWNTs sensitive layer

In this work we used ink-jet printing for DWNTs-based inks deposition. This friendly technique is low-cost and also known to be easily reproducible compared to other techniques used for CNTs deposition as drop casting for example. The main advantage of this technique is the dispensing and precise positioning of very small volumes of fluid (1–100 pL). With this approach narrow lines down to 20-25 um can be obtained, with a resolution of 5 μm, printed on all kinds of surface (rigid or flexible substrate) and with the ability to use ink solution having different properties. In this task, a Dimatix Materials Printer (Model DMP-2800, FUJIFILM Dimatix, Inc. Santa Clara, CA) was used as printer [9]. For ink-jet printing, homogenous ink with a good dispersion of the nanotubes is a crucial issue. In this study, DWNTs solution inks were prepared by sonicating oxidized CNTs in DMF solvent. We used a combination of tip sonication and shear-mixing. Typically, the suspensions are finally centrifuged in order to eliminate agglomerates and keep only the stable suspension in the supernatant. In the case of oxidized carbon nanotubes, this oxidation was performed by refluxing the DWNTs in a mixture of concentrated sulphuric and nitric acids at 70°C for 24h. After thorough washing to neutralization, the sample formed a very stable suspension in DMF [10].

Before printing on the $SiO_2$ surface, a surface treatment is needed in order to obtain a hydrophilic surface. The surface treatment used is $O_2$-plasma, with careful control in order to prevent strong effect on the guiding layer porosity, which may impact the mechanical wave propagation, subsequently implying a lot of insertion losses. Furthermore, the effect of the surface treatment is not permanent. After investigations, the power and duration were optimized: 30W $O_2$-plasma during 30s has been defined. With these parameters, the sample becomes properly hydrophilic for about 2 hours. The sample's temperature during the jetting was maintained at 80°C for adequate solvent evaporation.

On figure 1, a Love Wave sensor based on DWNTs is represented, which was selected for the SEM characterization of the DWNTs sensitive surface after printing. SEM analysis further confirmed that the DWNTs have clearly a high organization and an interesting surface morphology. Indeed, the huge surface to volume ratio will confer them outstanding sensitivity towards the molecules or particles absorbed on their surface.

### C. Electrical characterization

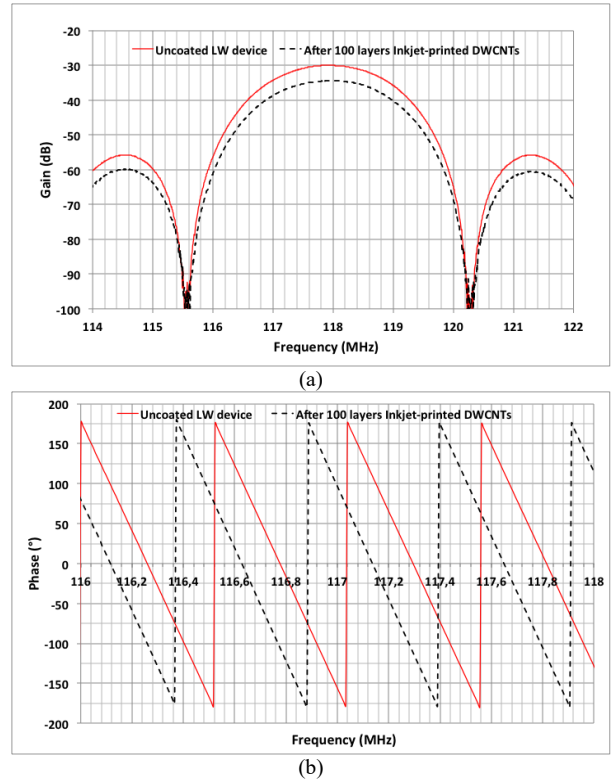

Fig. 2. (a) Gain and (b) phase, before and after inkjet - printed 100 layers of DWNTs as sensitive material (sample R31C4).

Electrical characterization (S parameters) of the platform was performed. A wide high-accuracy spectrum has been selected on the radio-frequency network analyzer to determine the minimum of insertion loss, of both the uncoated and coated delay lines.

The good performance characteristics of the coated sensor are illustrated on the figure 2, on which are represented the transmission line characteristics (S21) in terms of the gain (figure 2.a) and the phase (figure 2.b).

The sensor response was recorded before and after the deposition of the DWNTs inkjet-printed solutions. The working frequency was observed around 118 MHz. The induced insertion losses correspond to 4dB and the equiphase frequency shift was 160 kHz. Electrical characterization put to evidence that even high numbers of DWNTS layers, as high as 100 which corresponds to a thickness of 1.8um, can be deposited on the top of the Love wave sensor, while maintaining limited additional losses. Then, most of the acoustic wave energy will propagate in the DWNTs sensing layer, and thus the sensitivity will be significantly increased due to surface perturbations.

## III. RESULTS AND DISCUSSION

In this part of experiment, we study the behavior of this sensor based on DWCNTs sensitive material with ethanol and toluene vapors as well as with relative humidity.

The generation of vapors has been managed with a gas generator (Calibrage, PUL 110). A constant flow rate of Nitrogen, as a carrier gas (0.112 L/min) with a conventional sequence of different vapors concentrations circulated directly on both delay lines of the acoustic path, which were mounted into a specific hermetic cell. The concentration range was from 30 ppm to 500 ppm for VOCs vapors and from 2.11%HR to 41.92%HR for humidity.

Based on the real-time responses of the DWNTs-Love wave sensor, the sensitivities are extrapolated experimentally as:

$$S = \Delta f / \Delta C \qquad (1)$$

where $\Delta f$ is the frequency shift and $\Delta C$ represents the difference between the initial and the concentration of the target species. More details on the methodology followed for the generation of different concentrations of target spices in a carrier gas and the evaluation of the frequency shift are given in [11].

### A. VOCs detection

As represented on the figures 3 and 4, the real time detection of ethanol and toluene vapors shows clearly the evolution of the frequency shift as a function of the concentration levels of the target analytes. The influence of two different and subsequent exposures ($\Delta C$) is directly correlated with the frequency shift $\Delta f$, which is mainly based on the deviation from the initial concentration (30 ppm).

As shown on figure 5, the experimental sensitivity values correspond to 14.8 Hz/ppm and 13 Hz/ppm, for toluene and ethanol vapors, respectively. For example, this study under ethanol has put to evidence that printed DWNTs exhibit sensitivities higher than alternate functional materials studied in literature, such as $TiO_2$ (1.2 Hz/ppm) and Molecular Imprinted Polymer (MIP) (1.7 Hz/ppm) [12, 13].

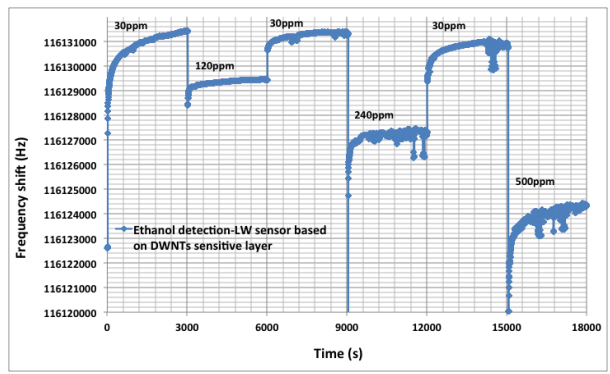

Fig. 3. Real-time detection of different concentrations of ethanol vapors.

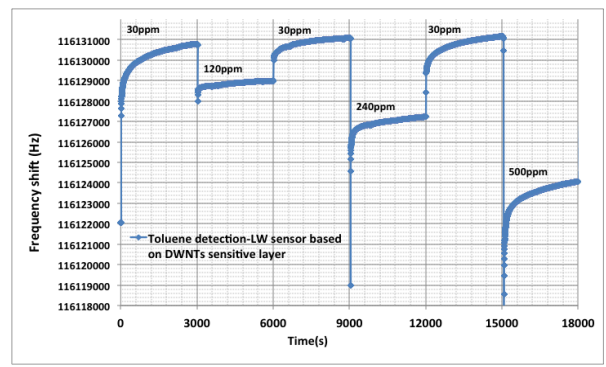

Fig. 4. Real-time detection of different concentrations of toluene vapors.

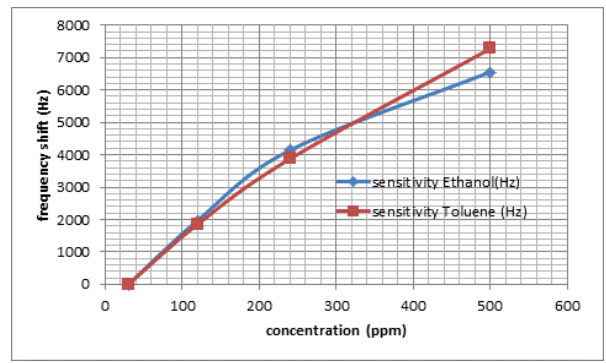

Fig. 5. Steady-state frequency shift of the sensor for the detection of ethanol and toluene vapors.

### B. Humidity detection

Similar experiments with humidity are represented on figure 6 and 7, which have shown a frequency shift of 2.93 kHz to 6.22 % RH and a quite linear sensitivity overall the tested range [2-42% RH].

From the steady-state frequency shift as a function of the concentration, as reported on figure 7, it can be estimated a sensitivity of 750 Hz/%RH.

Therefore, the results of detection of the ethanol and toluene vapors and humidity as shown by the recorded measurements demonstrate the potential of the Love wave sensor associated with DWNTs layers as a sensitive material.

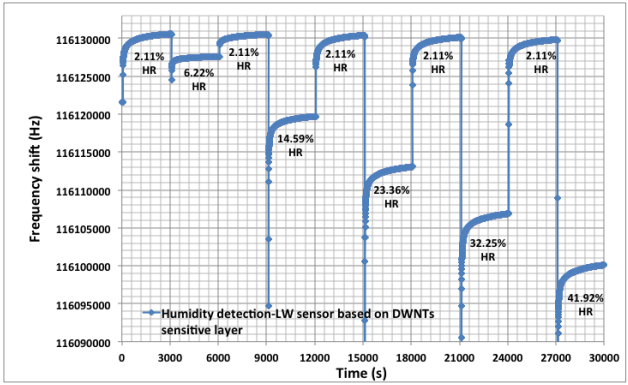

Fig. 6. Real-time detection of different concentrations of humidity.

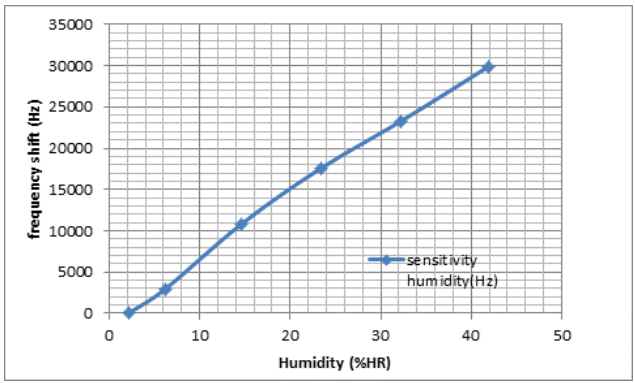

Fig. 7. Steady-state frequency shift of the sensor for the detection of humidity.

## IV. Conclusion

In summary, we reported the inkjet printing of DWNTs on Love wave platform and the fabrication of such DWNTs-based gas sensors. The electrical characterizations show that the printing of 100 layers of DWNTs introduce very low additional acoustic loss (- 4dB), which demonstrates the interest of the mechanical properties of this sensitive material.

The sensitivities of the sensor studied at different concentrations of vapors of ethanol, toluene and moisture at room temperature, were experimentally estimated to be 13 Hz/ppm, 14.8 Hz/ppm and 750 Hz/% RH, respectively.

As perspective, further study will focus on the Ink-jet printed DWNTS optimization in order to print smaller thickness of DWNTs (<1.8um), which should give faster response. In addition, new experiments will focus on functionalization of DWNTs regarding the selectivity of particular vapor detection. Thus, DWNTs-coated devices offer a real potential for application in air pollution meteorology and in industrial processes.

## Acknowledgements

Authors are grateful to the French National Research Agency (ANR-13-BS03-0010), the French RENATECH network (French National Nanofabrication Platform), and the French Embassy of Singapore (Merlion project).